\documentclass[final,letterpaper,5p]{elsarticle}
\usepackage{lineno}
\usepackage{graphicx,color}
\usepackage{amssymb}
\usepackage{amsmath}
\usepackage{latexsym}
\usepackage{amsxtra}
\usepackage{amscd}
\usepackage{dcolumn}% Align table columns on decimal point
\usepackage{bm}% bold math
\usepackage{hyperref}
\biboptions{numbers,sort&compress}
\let\oldsqrt\sqrt % it defines the new \sqrt in terms of the old one
\def\sqrt{\mathpalette\DHLhksqrt}
\def\DHLhksqrt#1#2{\setbox0=\hbox{$#1\oldsqrt{#2\,}$}\dimen0=\ht0
\advance\dimen0-0.2\ht0
\setbox2=\hbox{\vrule height\ht0 depth -\dimen0}%
{\box0\lower0.4pt\box2}}
\newcommand{\nuc}[2]{$^{#1}$#2}
\journal{Physics Letters B}

\begin{document}
\begin{frontmatter}
  
\title{Discovery of \nuc{68}{Br} in secondary reactions of radioactive beams}
\author[ut,rnc]{K.~Wimmer}
\author[rnc]{P.~Doornenbal}
\author[cea]{W.~Korten} 
\author[cce]{P.~Aguilera} 
\author[val,has]{A.~Algora}
\author[ut]{T.~Ando} 
\author[gsi,jlu]{T.~Arici} 
\author[rnc]{H.~Baba} 
\author[bor]{B.~Blank} 
\author[pad]{A.~Boso}
\author[rnc]{S.~Chen} 
\author[cea]{A.~Corsi} 
\author[uy]{P.~Davies} 
\author[leg]{G.~de Angelis} 
\author[gan]{G.~de France} 
\author[cea]{D.~T.~Doherty}  
\author[gsi]{J.~Gerl} 
\author[tum]{R.~Gernh\"{a}user} 
\author[uy]{D.~Jenkins} 
\author[ut]{S.~Koyama} 
\author[rnc]{T.~Motobayashi} 
\author[ut]{S.~Nagamine} 
\author[ut]{M.~Niikura} 
\author[cea,rnc]{A.~Obertelli} 
\author[tum]{D.~Lubos} 
\author[val]{B.~Rubio} 
\author[uo]{E.~Sahin} 
\author[ut]{T.~Y.~Saito} 
\author[ut,rnc]{H.~Sakurai} 
\author[uy]{L.~Sinclair} 
\author[rnc]{D.~Steppenbeck} 
\author[ut]{R.~Taniuchi} 
\author[uy]{R.~Wadsworth} 
\author[cea]{M.~Zielinska}

\address[ut]{Department of Physics, The University of Tokyo, 7-3-1 Hongo, Bunkyo-ku, Tokyo 113-0033, Japan}
\address[rnc]{RIKEN Nishina Center, 2-1 Hirosawa, Wako, Saitama 351-0198, Japan}
\address[cea]{IRFU, CEA, Universit\'{e} Paris-Saclay, F-91191 Gif-sur-Yvette, France}
\address[cce]{Comisi\'{o}n Chilena de Energ\'{i}a Nuclear, Casilla 188-D, Santiago, Chile}
\address[val]{Instituto de Fisica Corpuscular, CSIC-Universidad de Valencia, E-46071 Valencia, Spain}
\address[has]{Institute of Nuclear Research of the Hungarian Academy of Sciences, Debrecen H-4026, Hungary}
\address[gsi]{GSI Helmholtzzentrum f\"{u}r Schwerionenforschung, D-64291 Darmstadt, Germany}
\address[jlu]{Justus-Liebig-Universit\"{a}t Giessen, D-35392 Giessen, Germany}
\address[bor]{CENBG, CNRS/IN2P3, Universit\'{e} de Bordeaux F-33175 Gradignan, France}
\address[pad]{Istituto Nazionale di Fisica Nucleare, Sezione di Padova, I-35131 Padova, Italy}
\address[uy]{Department of Physics, University of York, YO10 5DD York, United Kingdom}
\address[leg]{Istituto Nazionale di Fisica Nucleare, Laboratori Nazionali di Legnaro, I-35020 Legnaro, Italy}
\address[gan]{GANIL, CEA/DSM-CNRS/IN2P3, F-14076 Caen Cedex 05, France}
\address[tum]{Physik Department, Technische Universit\"{a}t M\"{u}nchen, D-85748 Garching, Germany}
\address[uo]{Department of Physics, University of Oslo, PO Box 1048 Blindern, N-0316 Oslo, Norway}

\begin{abstract}
  The proton-rich isotope \nuc{68}{Br} was discovered in secondary fragmentation reactions of fast radioactive beams. Proton-rich secondary beams of \nuc{70,71,72}{Kr} and \nuc{70}{Br}, produced at the RIKEN Nishina Center and identified by the BigRIPS fragment separator, impinged on a secondary \nuc{9}{Be} target. Unambiguous particle identification behind the secondary target was achieved with the ZeroDegree spectrometer. Based on the expected direct production cross sections from neighboring isotopes, the lifetime of the ground or long-lived isomeric state of \nuc{68}{Br} was estimated. The results suggest that secondary fragmentation reactions, where relatively few nucleons are removed from the projectile, offer an alternative way to search for new isotopes, as these reactions populate preferentially low-lying states.
\end{abstract}

\date{\today}
\begin{keyword}
  radioactive beams, new isotope, direct reaction
\end{keyword}
\end{frontmatter}

The binding energy of a nucleus is a result of the interactions of all proton and neutron constituents. At the limits of nuclear binding, the neutron and proton drip lines, the separation energies become zero. A huge discovery potential is provided by the predicted 6900(500) bound nuclei~\cite{erler12}, compared with 3252 discovered to date~\cite{thoennessen13,thoennessen18}. In recent years, new isotopes have been mainly discovered by projectile fragmentation or fission of high intensity primary beams~\cite{hinke12,kurcewicz12,tarasov13,celikovic16,suzuki17,sumikama17,fukuda18,shimizu18}. Most notably, the discovery of \nuc{60}{Ca}, with $Z=20$ and $N=40$ a doubly closed Harmonic Oscillator shell nucleus, at the RIKEN Nishina Center~\cite{tarasov18} demonstrated the power of this technique.

On the proton-rich side, the limits of binding are experimentally established up to $A\sim80$, as the repulsive Coulomb force between the excess protons puts it much closer to the valley of stability. The Coulomb interaction, combined with the angular momentum barrier, can lead to long lifetimes of nuclei beyond the drip line, i.e., with negative proton separation energies. Such resonances can be reconstructed from the invariant mass measurement of the decay products. In some cases, these unbound nuclei are even sufficiently long lived to be measured as beam particles following projectile fragmentation and uniquely identified by measurements of $Z$ and $A$. For instance, the odd-odd isotope \nuc{72}{Rb} has been discovered and found to possess a half-life of 103(22)~ns~\cite{suzuki17}, whereas the less exotic \nuc{73}{Rb} is particle-unbound and unobserved with an upper limit for its half-life of 30~ns~\cite{pfaff96}. This fascinating observation of an odd-odd $Z>N$ isotope being longer-lived than its odd-even, less exotic neighboring isotope triggers the question whether this is an exceptional case or an indication of the stabilizing effect of the proton-neutron interaction.
% similar phenomena can occur in other regions of the nuclear chart.

In the vicinity of \nuc{72}{Rb}, the next heaviest case where the proton-neutron interaction could stabilize an odd-odd nucleus against immediate proton emission is \nuc{80}{Nb}, as \nuc{81}{Nb} does not exist or is extremely short-lived~\cite{janas99}. For even heavier nuclei, the drip-line is not well established. In lighter nuclei, several candidates for a more stable odd-odd nucleus than their less exotic neighboring isotope exist. As suggested for \nuc{72}{Rb}, the existence of such bridge nuclei may have implications on the nucleo-synthesis in the rapid proton capture process ($rp$-process)~\cite{woosley76}. Particularly the isotope \nuc{68}{Se}, with a half-life of 35.5~s a waiting-point for the $rp$-process, is of high significance, as the proton capture reaction on \nuc{68}{Se} would lead to the proton-unbound nucleus \nuc{69}{Br}. Two-proton capture on \nuc{68}{Se} allows this waiting point to be bypassed but the reaction rate depends exponentially on the proton separation energy of \nuc{69}{Br}. Alternatively, the $rp$-process could bypass the \nuc{68}{Se} waiting point through sequential proton capture on the lighter selenium isotope \nuc{67}{Se}, leading to an intermediate \nuc{68}{Br} and finally to \nuc{69}{Kr}. This scenario, however, requires the nucleus \nuc{68}{Br} to be bound, or at least sufficiently long-lived for proton capture.

In a search for new isotopes, proton-rich $N=33$ nuclei were produced at the NSCL~\cite{delsanto14}, and \nuc{68}{Br} remained unobserved suggesting that the isotope is unbound with a lifetime significantly shorter than the flight time through the separator. Also theoretical calculations as well as extrapolations from measured masses and $Q$-values predict that \nuc{68}{Br} is proton-unbound. Global mass predictions using the finite range droplet model predict $S_\text{p} = -110$~keV~\cite{moeller16}, while Hartree-Fock based calculations of Coulomb displacement energies result in $S_\text{p} = -710$~keV~\cite{brown02}. The latest atomic mass evaluation extrapolates the proton separation energy of \nuc{68}{Br} to $-500(250)$~keV~\cite{ame16}. Based on the above separation energy predictions, the lifetime estimates for \nuc{68}{Br} range from seconds to picoseconds when a barrier penetration model with the Wentzel-Kramers-Brillouin (WKB) approximation is applied. The shorter values are consistent with the non-observation after the flight time through the A1900 fragment separator of $\approx 440$ ns~\cite{delsanto14}. Also the most recent search for new neutron-deficient isotopes, carried out at the RIBF, identified the $N=29,31,32$ isotopes \nuc{63}{Se} and \nuc{67,68}{Kr}, but no event of \nuc{68}{Br} was observed~\cite{blank16}. As the fragment separator setting of BigRIPS~\cite{kubo12} was centered on the $N=30$ isotopes \nuc{65}{Br} and \nuc{64}{Se}, the large acceptance allowed only for partial transmission of $N=33$ isotones, and no limit on the lifetime of \nuc{68}{Br} could be established. From the non-observation of \nuc{68}{Br} in an earlier measurement performed at GANIL~\cite{blank95}, an upper limit of $\tau < 325$~ns could be deduced.
%as seen from the large numbers of identified \nuc{68}{Kr} and \nuc{66}{Se}. Due to the long flight time from the production target to the $Z$ identification, a lower limit of XYZ~ns can be inferred. %(I don't know if it's important to mention here a lower number for lifetime from Blank et al.) 

%{\it TODO: estimate half-life based on predicted separation energies}

%mass predictions FRDM~\cite{moeller16} $S_\text{p} = -622$~keV, AME16~\cite{ame16}  $S_\text{p} = -500(250)$~keV, Brown~\cite{brown02} $S_\text{p} = -710$~keV

In this Letter, we report on the first observation of the isotope \nuc{68}{Br} produced in secondary reactions of radioactive beams and provide a new lower lifetime limit for \nuc{69}{Br}. For the latter nucleus, initially claimed observations~\cite{mohar91} could not be confirmed later~\cite{blank95,pfaff96}. 

\begin{figure}[h]
\centering
\includegraphics[width=\columnwidth]{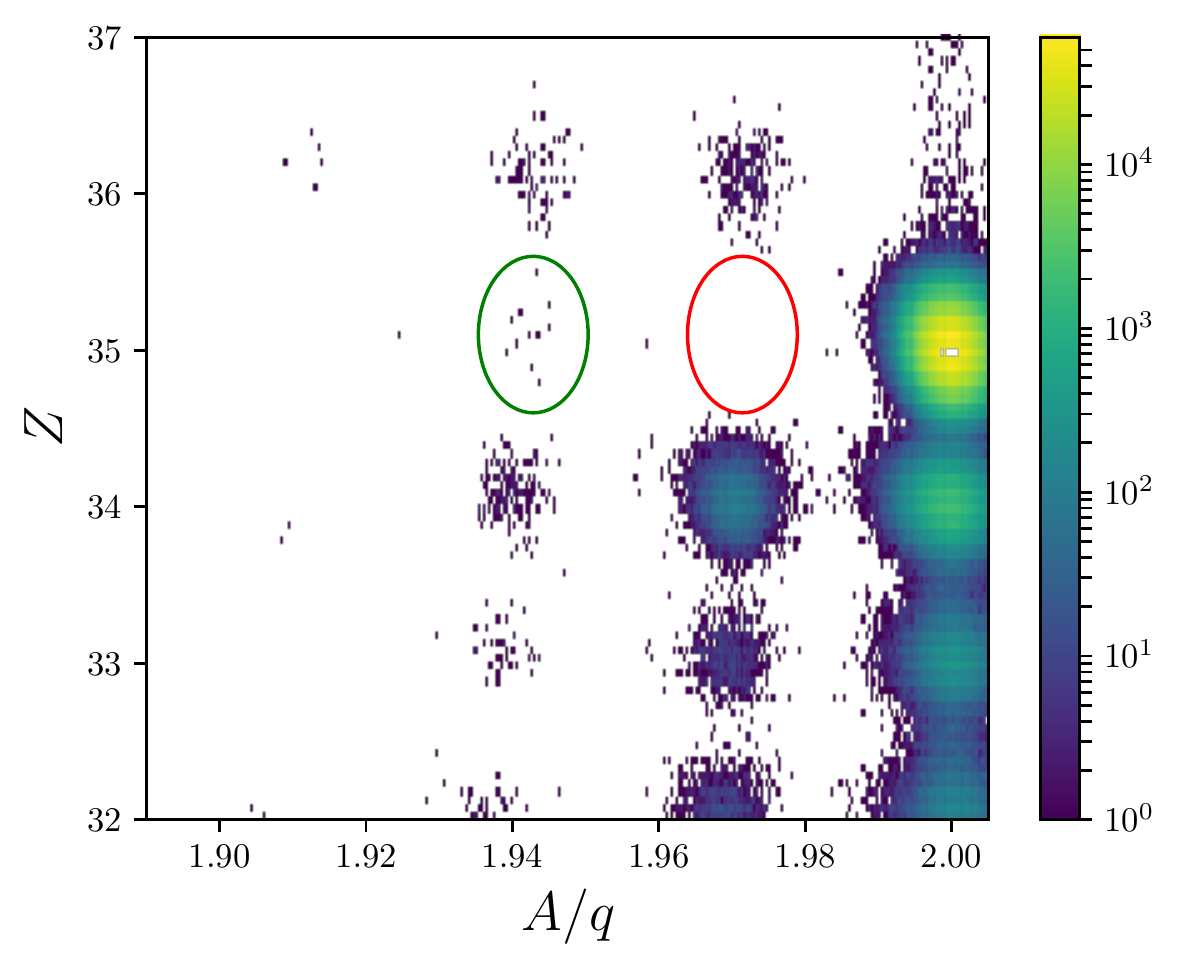}
\caption{Particle identification plot for the ZeroDegree spectrometer for \nuc{70}{Br} impinging on the \nuc{9}{Be} secondary reaction target. \nuc{68}{Br} is observed at $Z=35$, $A/q=1.943$ (green ellipse), \nuc{69}{Br} is absent (red ellipse).}
\label{fig:pid}
\end{figure}

The experiment has been performed at the Radioactive Isotope Beam Factory, operated by RIKEN Nishina Center and CNS, University of Tokyo. Radioactive nuclei in the vicinity of the $N=Z$ line were produced by projectile fragmentation of up to 250 pnA of \nuc{78}{Kr} primary beam at 345 MeV/u impinging on a 5~mm thick Be target. Secondary beams were purified and analyzed using the $B\rho-\Delta E-\text{TOF}$ technique in the BigRIPS separator~\cite{kubo12}. The flight time of the secondary beam through the separator amounted to $\sim450$~ns.
The fragmented secondary beams, which contained the isotopes \nuc{70}{Br}, \nuc{71}{Kr}, and \nuc{72}{Kr} at energies around 170~$A$MeV, impinged on a 703(7)~mg/cm$^2$ secondary Be reaction target located at the F8 focal point of the Big\-RIPS beam line. Knockout and few nucleon removal reaction products were analyzed in the ZeroDegree spectrometer~\cite{kubo12}. The flight time through this spectrometer amounted to $\sim270$~ns. Unique particle identification was achieved by measurements of energy loss, time-of-flight (TOF), and $B\rho$.
Fig.~\ref{fig:pid} shows the particle identification plot for the ZeroDegree spectrometer following \nuc{70}{Br} impinging on the secondary reaction target.
The proton-unbound nucleus \nuc{69}{Br}, indicated by the red ellipse for its expected location, is clearly not observed. This non-observation is in agreement with the upper limit on its half-life of 24~ns~\cite{pfaff96}. 
\begin{table*}[h]                 % optional [t, b or h];
  \caption{Number of observed events in the region of interest (ROI) of the particle identification plot, estimated background events, and corrected yields for the new isotope \nuc{68}{Br}.
    The inclusive lifetime is obtained from the yield at F11 compared to the expected yield calculated from the average -xpyn reaction cross sections of the same type. The uncertainties include the statistical uncertainties, the systematical uncertainties arising from the transmission through the ZeroDegree spectrometers, and the fluctuations of the cross sections for different isotopes as shown in Fig.~\ref{fig:averagecs}.}
  \begin{center}
    \label{tab:cs}
    {\renewcommand{\arraystretch}{1.1}
      \begin{tabular}{rrrrrrrr}
        \hline
        beam         & reaction & events in ROI & background & corrected yield   & $\sigma^{-xpyn}(\text{ave.)}$ (mb)  & expected yield & $\langle \tau \rangle$ (ns) \\% $\sigma$ ($\mu$b)      
        \hline                                                                               %                        
        \nuc{70}{Br} & -2n      &  12           & 0.15(2)    & 14.7(50)(18)      & 0.57(21)                            &  1740(710)     & 51(6)                       \\%  4.8(16)(14)           %  &  56(6)   \\ %0.0048(16)(14)\\
        \nuc{70}{Kr} & -1p1n    & 140           & 113(3)     & 33(16)(4)         & 82(12)                              &  2240(500)     & 57(7)                       \\% 1.0(5)(2)$\cdot 10^3$  %  & 61(7)    \\ %0.12(7)(1)\\
        \nuc{71}{Kr} & -1p2n    &  20           & 9.0(3)     & 13.7(60)(12)      & 5.7(12)                             &  2690(720)     & 46(6)                       \\% 31(14)(6)              %  &  51(9)   \\ %0.12(7)(1)\\
        \nuc{72}{Kr} & -1p3n    &  12           & 3.8(2)     & 10(4)(2)          & 0.32(8)                             &  1130(340)     & 51(6)                       \\%  2.8(13)(5)            %  &   59(10) \\ %0.0038(13)(11) \\
        \hline    
      \end{tabular}
    }
  \end{center}
\end{table*}
In contrast, 12 events are observed at $Z=35$, $A/q=1.943$ (green ellipse), corresponding to \nuc{68}{Br}. After correction for the detection efficiency and the transmission through the spectrometers, the yield of \nuc{68}{Br} amounts to 14.7(50)(18) with statistical and systematic uncertainties, respectively. The systematic uncertainties include the target thickness, detection efficiency, and transmission of the ZeroDegree spectrometer. Furthermore, 20 (12) events of \nuc{68}{Br} have also been observed following the secondary fragmentation of \nuc{71}{Kr} (\nuc{72}{Kr}) measured within the same experimental settings for BigRIPS and ZeroDegree. In the latter cases, however, background events of misidentified Kr nuclei have to be subtracted. Such events are caused by contamination from \nuc{71}{Kr} events undergoing reactions in the beam line detectors, especially in the plastic TOF stop detector and the ionization chamber at the final F11 focal point. Note that for these events the atomic charge $Z$ determined from the energy loss in the ionization chamber located behind the plastic scintillator can be lower. This leads to tails in the $Z$ identification. For \nuc{68}{Br}, the background can be estimated from comparison to the ratio of \nuc{71}{Kr} events and falsely observed \nuc{69}{Br}. Such a comparison leads to about 0.2~\% of the \nuc{70}{Kr} events creating a signal in the \nuc{68}{Br} region. This is in agreement with the estimated reaction rate using LISE++~\cite{tarasov08} simulations.
For the cleanest case of the \nuc{70}{Br} beam shown in Fig.~\ref{fig:pid}, the estimated background in the region of interest amounts to 0.15(2) events, while 12 have been observed. The estimated background for the different reaction channels is summarized in Table~\ref{tab:cs} together with observed events and transmission corrected yields.

\begin{figure}[h]
\centering
\includegraphics[width=\columnwidth]{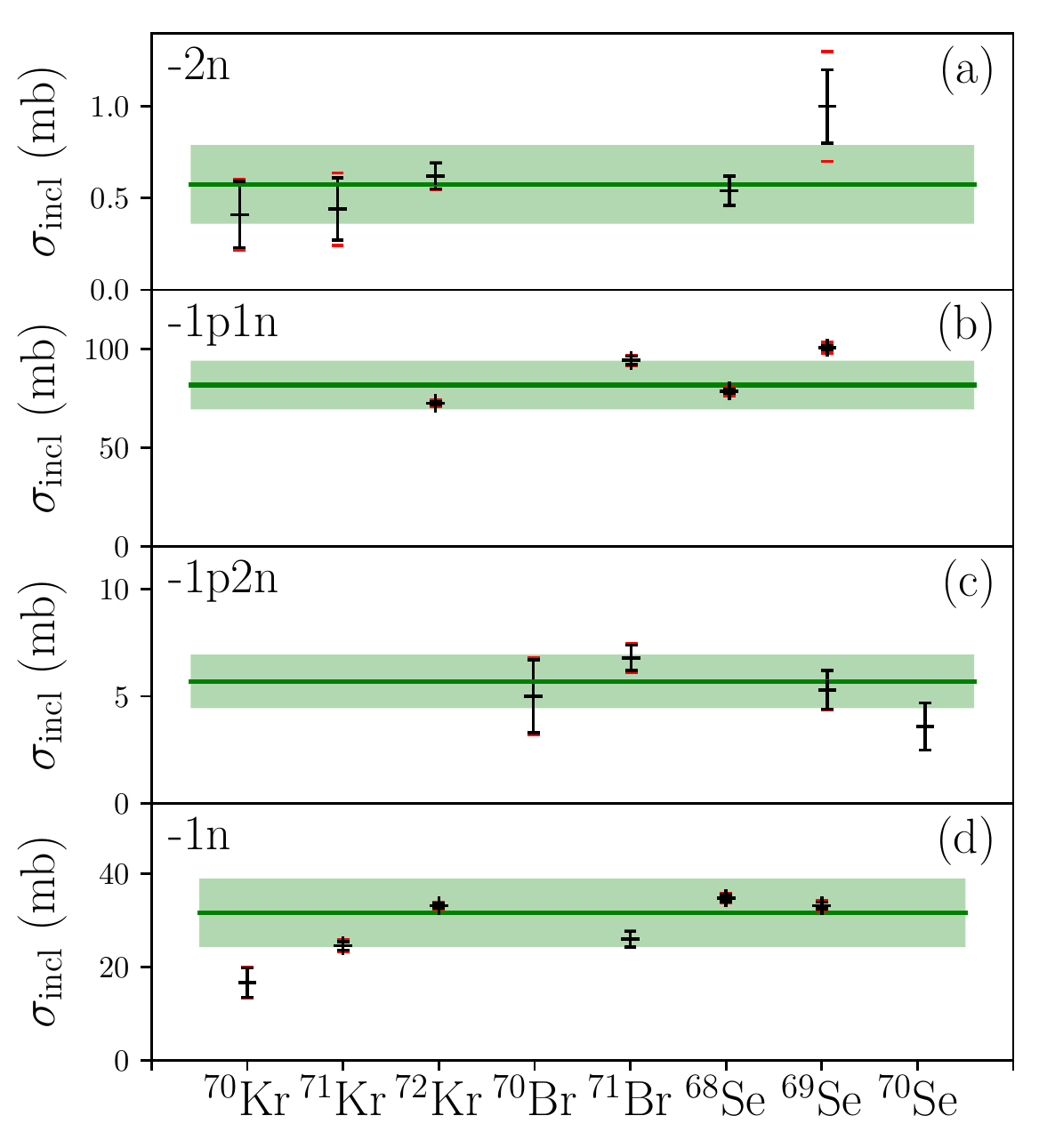}
\caption{Cross sections for various nucleon removal reactions. The x-axis is labeled with the projectile symbol. The statistical (black) and total (red) uncertainties have been taken into account to calculate the weighted cross sections for a certain reaction type. The green band shows the average and the fluctuations (standard deviation) which is used as an estimate for the uncertainty of the reaction cross section to \nuc{68}{Br}.}
\label{fig:averagecs}
\end{figure}

The measured yield is affected by the decay of \nuc{68}{Br} along the path to the final focal plane. In order to estimate the inclusive lifetime with respect to the state populated in the secondary fragmentation reaction (particle bound or unbound), the number of produced \nuc{68}{Br} ions needs to be estimated. Besides the reactions to \nuc{68}{Br}, the analogue -xpyn removal reactions have been measured in the same experiment for various isotopes in the same mass region. These reaction cross sections vary little between different isotopes. These reaction cross sections are shown in Fig.~\ref{fig:averagecs}.
The weighted average of the cross sections $\sigma^{-xpyn}(\text{ave.})$ for various -2n, -1p1n, and -1p2n reactions have been calculated to estimate the production of \nuc{68}{Br} at the secondary target. In the case of the -1p3n reaction, only one other case besides the \nuc{9}{Be}(\nuc{72}{Kr},X)\nuc{68}{Br} reaction was within the acceptance of ZeroDegree. The cross section for the \nuc{9}{Be}(\nuc{71}{Br},X)\nuc{67}{Se} reaction amounts to $\sigma = 0.32(6)(5)$~mb.
These cross sections and the corresponding expected yields $Y^{-xpyn}(\text{exp.})$ are also given in Table~\ref{tab:cs}.

The lifetime of \nuc{68}{Br} has then been estimated from the length of the flight path to the focal plane $T_\text{ZDS}$ using the relation including the relativistic $\gamma$ factor to account for the time dilation:
\begin{equation}
  \tau = \frac{1}{\gamma}\frac{t_\text{ZDS}}{\ln{(Y^{-xpyn}(\text{exp.}))}-\ln{(Y^{-xpyn}(^{68}\text{Br}))}}
  \label{eq:tau}
\end{equation} 
Estimated values, which range between 45--57~ns, are included in Table~\ref{tab:cs}. It should be noted that the uncertainties on the lifetime are rather insensitive to the average cross sections determined from Fig.~\ref{fig:averagecs}. A factor of two scaling of the number of expected counts results in a lifetime difference of less than 10~ns.
The expected counts listed in Table~\ref{tab:cs} are the final state inclusive ones, but fewer bound states are expected for \nuc{68}{Br} in comparison to the neighboring, less proton-rich, nuclei. For a few cases statistics were sufficient to analyze also the exclusive cross sections using the DALI2 $\gamma$-ray spectrometer~\cite{takeuchi14} (details are found in Ref~\cite{wimmer18}): The two-neutron removal reactions from \nuc{72}{Kr}, the -1p1n reactions from \nuc{72}{Kr} and \nuc{71}{Br}, as well as the -1p2n reaction from \nuc{71}{Br}. Analysis of these reaction channels revealed 65--90~\% ground state population (or very low-lying, unobserved, states below 200 keV in the case of the \nuc{9}{Be}(\nuc{71}{Br},X)\nuc{69}{Se} reaction), and are thus comparable to the available final states of \nuc{68}{Br}.

The production of \nuc{68}{Br} through few-nucleon removal reactions from neighboring isotopes populates preferentially low-lying low-spin states in the ejectiles, while in the direct fragmentation of \nuc{78}{Kr} population of very short-lived particle-unbound states at higher excitation energy are favored. Additionally, shorter flight times enhance the survival probability of secondary reaction products with very short lifetimes.
Both effects may explain the non-observation of \nuc{68}{Br} in the secondary beams of earlier searches for new isotopes. With an estimated lifetime around 50 ns, \nuc{68}{Br} could nevertheless have been observed also at the BigRIPS separator~\cite{blank16} or at the NSCL~\cite{delsanto14}. 
In the present experiment, a conservative upper limit for the experimental yield of \nuc{68}{Br} after the BigRIPS separator amounted to 200 counts, which is more than two orders of magnitude lower than for \nuc{70}{Kr}. With the EPAX3 parametrization~\cite{suemmerer12}, the calculated cross section for the production of \nuc{68}{Br} amounts to $\sigma = 1.29\cdot 10^{-5}$~mb, a factor of two higher than for \nuc{70}{Kr}.
This suggests that states of \nuc{68}{Br} produced in the fragmentation of the primary beam are short-lived, with lifetimes much shorter than the flight time through the separator, in line with the non-observation of \nuc{68}{Br} in earlier experiments~\cite{delsanto14,blank16}. The present experiment was not optimized for \nuc{68}{Br} and therefore the transmission of this isotope is not well determined. Therefore, an upper limit on the lifetime of \nuc{68}{Br} cannot be obtained from the primary beam fragmentation. 

The fragmentation of the primary \nuc{78}{Kr} beam is expected to produce \nuc{68}{Br} mainly through proton evaporation from the pre-fragment \nuc{69}{Kr}. In the Abrasion-Ablation model~\cite{gaimard91}, mean excitation energies of the pre-fragments around 100~MeV are predicted. Thus, the survival probability of exotic fragments around the drip-lines is rather low, even if the angular momentum barrier is taken into account. For the present case, the mean angular momentum is estimated~\cite{dejong97} at $\langle J\rangle~\sim 5\hbar$ with a width ($\sigma$) of $5\hbar$. This favors population of medium- and low-spin states. Conversely, the few nucleon removal reactions from radioactive secondary beams observed in this experiment predominantly populate the ground state directly. As a consequence --and in combination with the much lower flight time for identification--, the much lower secondary beam intensities were overcome, allowing the first observation of the very exotic nucleus \nuc{68}{Br}.

Similar arguments can also be used to estimate the production cross section of \nuc{69}{Br} through one-neutron knockout from the \nuc{70}{Br} nucleus. On average, the one-neutron knockout cross sections for $A\sim70$ nuclei amount to $32(7)$~mb, with some dependence on the separation energy when approaching the drip-line (see Fig.~\ref{fig:averagecs} (d)). Assuming this cross section for the \nuc{9}{Be}(\nuc{70}{Br},X)\nuc{69}{Br} reaction, on the order of $9\cdot10^4$ counts should have been observed in the particle identification plot shown in Fig.~\ref{fig:pid}. The non-observation leads to a 5$\sigma$ upper limit of 15 true events in the region of interest at \nuc{69}{Br}. With the lifetime estimate from Eq.~\ref{eq:tau} this leads to a very conservative upper limit of $\tau<28$~ns for the lifetime. It has been argued previously that the non-observation of \nuc{69}{Br} could be due to the preferred population of isomeric states, with shorter lifetimes than the ground state and a large particle decay width~\cite{jenkins08}. The non-observation in the present direct reaction, however, suggests that the ground state itself is unbound with a very short lifetime.

\begin{figure}[htb]
\centering 
\includegraphics[width=\columnwidth]{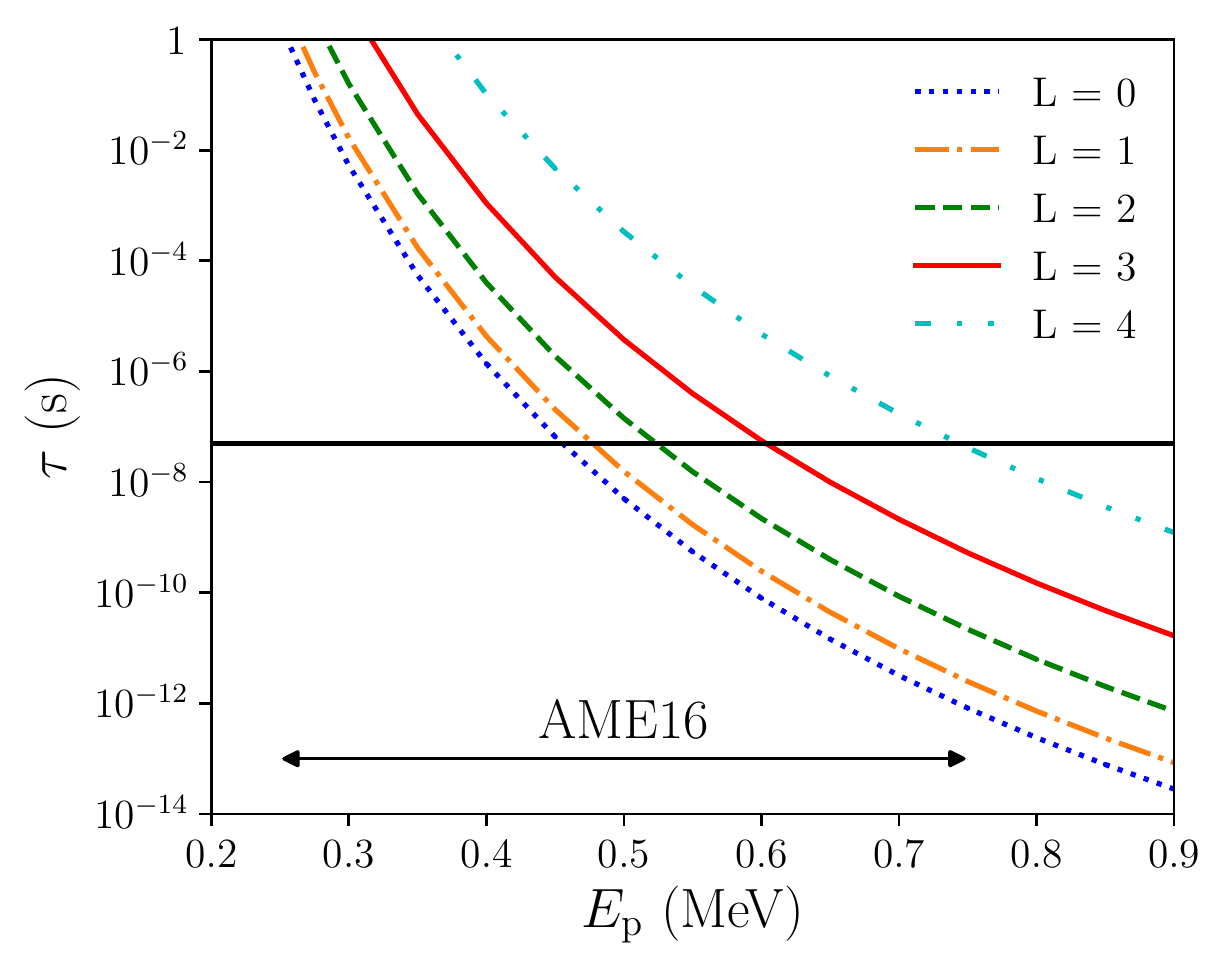}
\caption{Lifetime $\tau$ of \nuc{68}{Br} calculated using a barrier penetration model. For various values of the angular momentum transfer $L$ of the decay, the lifetime depends strongly on the separation energy $S_\text{p} =-E_p$. The extrapolation of the atomic mass evaluation amounts to $-500(250)$~keV~\cite{ame16}, indicated by the arrow.}
\label{fig:tau}
\end{figure}

Going one step further, the proton-separation energy can be estimated by applying a barrier penetration model with the WKB approximation. Assuming that the lifetime of $\tau\sim50$~ns corresponds to the ground state of \nuc{68}{Br}, its separation energy is calculated to be of the order of $S_\text{p} = -600$~keV. Fig.~\ref{fig:tau} shows these calculated lifetimes $\tau$ as a function of the energy $E_p$ (proton separation energy $S_\text{p}=-E_p$) for various values of the angular momentum of the proton decay $L$. Spherical Woods-Saxon type potentials were used to model the nuclear potential in these calculations.

The ground state spin and parity of \nuc{68}{Br} is unknown, but the one for the mirror nucleus, \nuc{68}{As}, is known to be $J^\pi=3^+$, and, using isospin symmetry arguments, the same can be assumed for \nuc{68}{Br}. The proton decay daughter \nuc{67}{Se} has a $J^\pi=5/2^-$ ground state and an almost degenerate low-lying $3/2^-$ state~\cite{orlandi09}. The emission of a $f_{5/2}$ proton with $L=3$ is consistent with the lifetime estimate for \nuc{68}{Br} and a proton separation energy of $S_\text{p} \approx -600$~keV. Conversely, decay to the $3/2^-$ state via the emission of a $L=1$ $p_{3/2}$ proton would suggest that \nuc{68}{Br} is more bound, with $S_\text{p} \approx -480$~keV. Unfortunately, the structure of \nuc{68}{Br} cannot be inferred from the present experiment and future studies of $\beta$-delayed proton emission from \nuc{68}{Kr} similar to the study of \nuc{69}{Br}~\cite{delsanto14} are required to clarify this situation. 

In summary, the isotope \nuc{68}{Br}, located beyond the proton drip-line, has been observed for the first time. In contrast to previous attempts to produce \nuc{68}{Br}, the production by few-nucleon removal reactions from radioactive secondary beams leads to the population of low-lying states. Comparison of the observed yield to similar reactions in the same mass region leads to a lifetime estimate of about 50~ns for \nuc{68}{Br}. Its existence can be explained by a stabilization through the proton-neutron interaction. The same feature might also occur in other odd-odd nuclei beyond the drip-line. \nuc{62}{As}, \nuc{58}{Ga}, \nuc{48}{Co}, \nuc{44}{Mn}, \nuc{38}{Sc}, and potentially others could be more stable than their less exotic even-odd neighboring isotopes. The production of nuclei at and beyond the proton drip-line in the rapid proton capture process can then influence its timescale through the potential by-pass of waiting points and the light curve or final abundance of nuclei. It is therefore important to study the stability of these nuclei, and, as presented in this study, the production by few-nucleon removal reactions instead of fusion evaporation or projectile fragmentation could be better suited as a discovery tool for nuclei beyond the proton drip-line.
We conclude with the remark that the observation of \nuc{68}{Br} through the secondary reactions and not in the fragmentation of the primary beam suggests that states populated through the fragmentation reaction are unbound and particle-decay before observation is possible. Conversely, few-nucleon removal reactions predominantly populate low-lying states. Therefore, lifetime estimates obtained from the observation or non-observation of isotopes produced by projectile fragmentation of primary beams might be off by orders of magnitude.

%\acknowledgments
We would like to thank the RIKEN accelerator and BigRIPS teams for providing the high intensity beams. 
This work has been supported by UK STFC under grant numbers ST/L005727/1 and ST/P003885/1, the Spanish Ministerio de Econom\'ia y Competitividad under grants FPA2011-24553 and FPA2014-52823-C2-1-P and the Program Severo Ochoa (SEV-2014-0398).
A.O. acknowledges the support from the European Research Council through the ERC Grant No. MINOS-258567.

\bibliographystyle{elsarticle-num-names}
\bibliography{br68resubmit}

\begin{thebibliography}{31}
\providecommand{\natexlab}[1]{#1}
\providecommand{\url}[1]{\texttt{#1}}
\providecommand{\urlprefix}{URL }
\expandafter\ifx\csname urlstyle\endcsname\relax
  \providecommand{\doi}[1]{doi:\discretionary{}{}{}#1}\else
  \providecommand{\doi}[1]{doi:\discretionary{}{}{}\begingroup
  \urlstyle{rm}\url{#1}\endgroup}\fi
\providecommand{\bibinfo}[2]{#2}

\bibitem[{Erler et~al.(2012)}]{erler12}
\bibinfo{author}{J.~Erler}, et~al., \bibinfo{journal}{Nature}
  \bibinfo{volume}{738} (\bibinfo{year}{2012}) \bibinfo{pages}{453}.

\bibitem[{Thoennessen(2013)}]{thoennessen13}
\bibinfo{author}{M.~Thoennessen}, \bibinfo{journal}{Reports on Progress in
  Physics} \bibinfo{volume}{76} (\bibinfo{year}{2013}) \bibinfo{pages}{056301}.

\bibitem[{Thoennessen(2018)}]{thoennessen18}
\bibinfo{author}{M.~Thoennessen}, \bibinfo{title}{Discovery of Nuclides
  Project},
  \bibinfo{howpublished}{\url{https://people.nscl.msu.edu/~thoennes/isotopes/}},
  \bibinfo{year}{2018}.

\bibitem[{Hinke et~al.(2012)}]{hinke12}
\bibinfo{author}{C.~Hinke}, et~al., \bibinfo{journal}{Nature}
  \bibinfo{volume}{341} (\bibinfo{year}{2012}) \bibinfo{pages}{486}.

\bibitem[{Kurcewicz et~al.(2012)}]{kurcewicz12}
\bibinfo{author}{J.~Kurcewicz}, et~al., \bibinfo{journal}{Phys. Lett. B}
  \bibinfo{volume}{717} (\bibinfo{year}{2012}) \bibinfo{pages}{371}.

\bibitem[{Tarasov et~al.(2013)}]{tarasov13}
\bibinfo{author}{O.~Tarasov}, et~al., \bibinfo{journal}{Phys. Rev. C}
  \bibinfo{volume}{87} (\bibinfo{year}{2013}) \bibinfo{pages}{054612}.

\bibitem[{Celikovic et~al.(2016)}]{celikovic16}
\bibinfo{author}{I.~Celikovic}, et~al., \bibinfo{journal}{Phys. Rev. Lett.}
  \bibinfo{volume}{116} (\bibinfo{year}{2016}) \bibinfo{pages}{162501}.

\bibitem[{Suzuki et~al.(2017)}]{suzuki17}
\bibinfo{author}{H.~Suzuki}, et~al., \bibinfo{journal}{Phys. Rev. Lett.}
  \bibinfo{volume}{119} (\bibinfo{year}{2017}) \bibinfo{pages}{192503}.

\bibitem[{Sumikama et~al.(2017)}]{sumikama17}
\bibinfo{author}{T.~Sumikama}, et~al., \bibinfo{journal}{Phys. Rev. C}
  \bibinfo{volume}{95} (\bibinfo{year}{2017}) \bibinfo{pages}{051601}.

\bibitem[{Fukuda et~al.(2018)}]{fukuda18}
\bibinfo{author}{N.~Fukuda}, et~al., \bibinfo{journal}{J. Phys. Soc. Jpn.}
  \bibinfo{volume}{87} (\bibinfo{year}{2018}) \bibinfo{pages}{014202}.

\bibitem[{Shimizu et~al.(2018)}]{shimizu18}
\bibinfo{author}{Y.~Shimizu}, et~al., \bibinfo{journal}{J. Phys. Soc. Jpn.}
  \bibinfo{volume}{87} (\bibinfo{year}{2018}) \bibinfo{pages}{014203}.

\bibitem[{Tarasov et~al.(2018)}]{tarasov18}
\bibinfo{author}{O.~B. Tarasov}, et~al., \bibinfo{journal}{Phys. Rev. Lett.}
  \bibinfo{volume}{121} (\bibinfo{year}{2018}) \bibinfo{pages}{022501}.

\bibitem[{Pfaff et~al.(1996)}]{pfaff96}
\bibinfo{author}{R.~Pfaff}, et~al., \bibinfo{journal}{Phys. Rev. C}
  \bibinfo{volume}{53} (\bibinfo{year}{1996}) \bibinfo{pages}{1753--1758}.

\bibitem[{Janas et~al.(1999)}]{janas99}
\bibinfo{author}{Z.~Janas}, et~al., \bibinfo{journal}{Phys. Rev. Lett.}
  \bibinfo{volume}{82} (\bibinfo{year}{1999}) \bibinfo{pages}{295--298}.

\bibitem[{Woosley and Taam(1976)}]{woosley76}
\bibinfo{author}{S.~E. Woosley}, \bibinfo{author}{R.~E. Taam},
  \bibinfo{journal}{Nature} \bibinfo{volume}{263} (\bibinfo{year}{1976})
  \bibinfo{pages}{101}.

\bibitem[{Santo et~al.(2014)}]{delsanto14}
\bibinfo{author}{M.~D. Santo}, et~al., \bibinfo{journal}{Physics Letters B}
  \bibinfo{volume}{738} (\bibinfo{year}{2014}) \bibinfo{pages}{453 -- 456}.

\bibitem[{M{\"o}ller et~al.(2016)M{\"o}ller, Sierk, Ichikawa, and
  Sagawa}]{moeller16}
\bibinfo{author}{P.~M{\"o}ller}, \bibinfo{author}{A.~Sierk},
  \bibinfo{author}{T.~Ichikawa}, \bibinfo{author}{H.~Sagawa},
  \bibinfo{journal}{Atomic Data and Nuclear Data Tables} \bibinfo{volume}{109}
  (\bibinfo{year}{2016}) \bibinfo{pages}{1}.

\bibitem[{Brown et~al.(2002)Brown, Clement, Schatz, Volya, and
  Richter}]{brown02}
\bibinfo{author}{B.~A. Brown}, \bibinfo{author}{R.~R.~C. Clement},
  \bibinfo{author}{H.~Schatz}, \bibinfo{author}{A.~Volya},
  \bibinfo{author}{W.~A. Richter}, \bibinfo{journal}{Phys. Rev. C}
  \bibinfo{volume}{65} (\bibinfo{year}{2002}) \bibinfo{pages}{045802}.

\bibitem[{Huang et~al.(2017)Huang, Audi, Wang, Kondev, Naimi, and Xu}]{ame16}
\bibinfo{author}{W.~J. Huang}, \bibinfo{author}{G.~Audi},
  \bibinfo{author}{M.~Wang}, \bibinfo{author}{F.~G. Kondev},
  \bibinfo{author}{S.~Naimi}, \bibinfo{author}{X.~Xu},
  \bibinfo{journal}{Chinese Physics C} \bibinfo{volume}{41}
  (\bibinfo{year}{2017}) \bibinfo{pages}{030003}.

\bibitem[{Blank et~al.(2016)}]{blank16}
\bibinfo{author}{B.~Blank}, et~al., \bibinfo{journal}{Phys. Rev. C}
  \bibinfo{volume}{93} (\bibinfo{year}{2016}) \bibinfo{pages}{061301}.

\bibitem[{Kubo(2012)}]{kubo12}
\bibinfo{author}{T.~Kubo}, \bibinfo{journal}{Prog. Theor. Exp. Phys.}
  \bibinfo{volume}{2012} (\bibinfo{year}{2012}) \bibinfo{pages}{03C003}.

\bibitem[{Blank et~al.(1995)}]{blank95}
\bibinfo{author}{B.~Blank}, et~al., \bibinfo{journal}{Phys. Rev. Lett.}
  \bibinfo{volume}{74} (\bibinfo{year}{1995}) \bibinfo{pages}{4611--4614}.

\bibitem[{Mohar et~al.(1991)}]{mohar91}
\bibinfo{author}{M.~F. Mohar}, et~al., \bibinfo{journal}{Phys. Rev. Lett.}
  \bibinfo{volume}{66} (\bibinfo{year}{1991}) \bibinfo{pages}{1571}.

\bibitem[{Tarasov and Bazin(2008)}]{tarasov08}
\bibinfo{author}{O.~Tarasov}, \bibinfo{author}{D.~Bazin},
  \bibinfo{journal}{Nucl. Instr. Meth. B} \bibinfo{volume}{266}
  (\bibinfo{year}{2008}) \bibinfo{pages}{4657}.

\bibitem[{Takeuchi et~al.(2014)}]{takeuchi14}
\bibinfo{author}{S.~Takeuchi}, et~al., \bibinfo{journal}{Nucl. Instr. Meth. A}
  \bibinfo{volume}{763} (\bibinfo{year}{2014}) \bibinfo{pages}{596}.

\bibitem[{Wimmer et~al.(2018)}]{wimmer18}
\bibinfo{author}{K.~Wimmer}, et~al., \bibinfo{journal}{Phys. Lett. B}
  \bibinfo{volume}{785} (\bibinfo{year}{2018}) \bibinfo{pages}{441}.

\bibitem[{S\"ummerer(2012)}]{suemmerer12}
\bibinfo{author}{K.~S\"ummerer}, \bibinfo{journal}{Phys. Rev. C}
  \bibinfo{volume}{86} (\bibinfo{year}{2012}) \bibinfo{pages}{014601}.

\bibitem[{Gaimard and Schmidt(1991)}]{gaimard91}
\bibinfo{author}{J.-J. Gaimard}, \bibinfo{author}{K.-H. Schmidt},
  \bibinfo{journal}{Nuclear Physics A} \bibinfo{volume}{531}
  (\bibinfo{year}{1991}) \bibinfo{pages}{709 -- 745}.

\bibitem[{de~Jong et~al.(1997)de~Jong, Ignatyuk, and Schmidt}]{dejong97}
\bibinfo{author}{M.~de~Jong}, \bibinfo{author}{A.~Ignatyuk},
  \bibinfo{author}{K.-H. Schmidt}, \bibinfo{journal}{Nuclear Physics A}
  \bibinfo{volume}{613} (\bibinfo{year}{1997}) \bibinfo{pages}{435 -- 444}.

\bibitem[{Jenkins(2008)}]{jenkins08}
\bibinfo{author}{D.~G. Jenkins}, \bibinfo{journal}{Phys. Rev. C}
  \bibinfo{volume}{78} (\bibinfo{year}{2008}) \bibinfo{pages}{012801}.

\bibitem[{Orlandi et~al.(2009)}]{orlandi09}
\bibinfo{author}{R.~Orlandi}, et~al., \bibinfo{journal}{Phys. Rev. Lett.}
  \bibinfo{volume}{103} (\bibinfo{year}{2009}) \bibinfo{pages}{052501}.

\end{thebibliography}

\end{document}